# Flat electronic band structure and anisotropic optical, mechanical, and thermoelectric properties of two-dimensional fullerene networks

Linfeng Yu[1], Jinyuan Xu[1], Bo Peng[1], Guangzhao Qin[1,*], and Gang Su[2,†]

[1]*State Key Laboratory of Advanced Design and Manufacturing for Vehicle Body, College of Mechanical and Vehicle Engineering, Hunan University, Changsha 410082, P. R. China*

[2]*Kavli Institute for Theoretical Sciences and CAS Center for Excellence in Topological Quantum Computation, University of Chinese Academy of Sciences, Beijing 100190, China*

*Correspondence:* gzqin@hnu.edu.cn, gsu@ucas.ac.cn

**Abstract**

Nanoclusters like fullerenes as the unit to build intriguing two-dimensional (2D) topological structures is of great challenge. Here we propose three bridged fullerene monolayers and comprehensively investigate the novel fullerene monolayer (α-$C_{60}$-2D) as synthesized experimentally [*Hou et al.*, *Nature* **606**, 507–510 (2022)] by s*tate-of-the-art* first-principles calculations. Our results show that α-$C_{60}$-2D has a direct bandgap of 1.49 eV owing to a flat conduction band bottom (CBM) close to the experimental value, the optical linear dichroism with strong absorption in long-wave ultraviolet region, a small anisotropic Young's modulus, the large hole mobility, and the ultrahigh Seebeck coefficient at middle-low temperatures. Moreover, Li ions are found to migrate easily along the X path in α-$C_{60}$-2D. It is unveiled that the anisotropic optical, mechanical, electrical, and thermoelectric properties of α-$C_{60}$-2D originate from the asymmetric bridging arrangements between $C_{60}$ clusters. Our study promises potential applications of monolayer fullerene networks in diverse fields.

**Keywords**: carbon, $C_{60}$, fullerene network

**INTRODUCTION**

Diverse bonding forms with $sp^n$ (n=1, 2, or 3) hybridizations lead to rich configurations of carbon (C) materials, which gives rise to zero- (0D), one- (1D), two- (2D) and three-dimensional (3D) materials, such as fullerenes[1,2], carbon nanotubes[3,4], graphene[5–7], diamond[8,9], *etc*. Through the reconstruction of the bonding form, carbon materials can physically bridge the barriers in metals, semimetals, semiconductors, and insulators[10–12], exhibiting rich and fascinating properties. Some novel carbon materials were obtained either in experiments or in theoretical predictions, such as 3D T-carbon[13], Cco-$C_8$[14], V-carbon[15], and 2D graphdiyne[16], biphenylene[10], octagraphene[17], graphene+[18], penta-graphene[19], and so on. So far, 524 carbon allotropes have been included in the SCADA[20] database, revealing the prosperity and diversity of carbon family in materials science.

2D carbon materials derived from graphene[5] exhibit nontrivial physical properties due to quantum effects, including ultra-high carrier mobility[6,21], superconductivity[22–24], and auxetic properties[18,19]. The current 2D carbon allotropes are mainly formed by periodic reconstruction of carbon elements based on the unit of single atoms, generating planar or buckling carbon phases. Promising structural motifs, like nanoclusters, were successfully applied to predict novel carbon allotropes such as T-carbon[13], which has been synthesized[25,26]. The 2D structures constructed by assembling nanoclusters are expected to possess exotic topology and non-trivial physical properties. Fullerenes, as 0D carbon clusters, can also be regularly arranged periodically and utilized to build layered polymers in form of covalent bonds between clusters, exhibiting interesting physical properties[27–30]. However, strong interlayer covalent bonds hinder the preparation of atomically-scale 2D $C_{60}$ monolayers from layered $C_{60}$ polymers due to the metastability[29]. Recently, *Zheng et al*. successfully fabricated a novel 2D $C_{60}$ monolayer by an interlayer bond cleavage strategy[31], demonstrating a new bridging assembly between $C_{60}$ clusters via covalent bonds.

In this paper, we theoretically study the novel 2D $C_{60}$ polymer monolayer with a quasi-hexagonal bridged form. The geometric and electrical properties of α-$C_{60}$-2D polymer monolayer are comprehensively explored based on first-principles calculations. The flat electronic band structure leads to dual conduction band minimum (CBM), making α-$C_{60}$-2D a direct-bandgap semiconductor. In addition, anisotropic optical, mechanical, and electrical transport properties are found in α-$C_{60}$-2D, which originates from the asymmetric bridge bonding arrangement. The promising properties would push forward a rapid advance of potential applications of the nanocluster-designed 2D carbon monolayers in micro- and nano-electronics.

**RESULTS AND DISCUSSION**

**Geometrical structures** -. Fullerenes can be connected to construct a network via different ways. Figs. 1(a–d) present three assembled configurations, forming $C_{60}$ polymer monolayers (α-$C_{60}$, β-$C_{60}$, and γ-$C_{60}$) and their corresponding van der Waals structures (vdW-3D). As shown in Fig. 1(a), α-$C_{60}$-2D is assembled by directly bridging covalently bonded six fullerenes at ~60° intervals along another fullerene, where the C atoms of two adjacent fullerenes are connected by single $sp^3$ bond along ~60, 120, 240, and 300°, while the C atoms are connected by two $sp^3$ bonds along 0, and 180° directions. The formation of fullerene monolayers of β- [Fig. 1(b)] and γ-phase [Fig. 1(c)] is achieved by bridging four fullerenes along the orthogonal direction of the central fullerene, respectively. Lattice constants and more details can be found in the Supplemental Table S1. The different assembly configurations lead to the diversity of physical properties.

To compare with experiments, we carried out the single-crystal X-ray diffraction (XRD) simulations as shown in Fig. 1(e). It is found that the α-phase layered $C_{60}$ polymer (α-$C_{60}$-3D) exhibits a strong diffraction peak at $2\theta = 8.5°$, which corresponds well to the experimentally strong diffraction peak at $2\theta = 8.8°$. The two $sp^3$-hybridized bonds connect adjacent fullerene spheres along the parallel directions, while single bonds are in the orthogonal direction. The characteristics indicate that the intercluster Mg-C of the α-phase $C_{60}$ polymer prepared by *Zheng et al*. is cleaved by organic cation exchange to form the $C_{60}$ polymer with vdW interaction[31]. Such a vdW phase can be well decomposed into the atomic-level $C_{60}$ cluster monolayer with stable and ordered configuration[32].

**Stability and bridge bond** -. The α-$C_{60}$ prepared based on the interlayer cracking strategy is thermodynamically stable, which can be verified from different aspects. As shown in Fig. 1(f), the energies and corresponding vdW configurations of all fullerene monolayers are higher than those of graphene[11], diamond[8], and Bct-$C_4$[33], but still lower than graphene+[18], T-carbon[13], penta-graphene[19], demonstrating the energetic metastability. Furthermore, the micro-canonical ensemble (NVE) was first implemented to inspect the thermal stability at different temperatures based on *ab-initio* molecular dynamics (AIMD) simulations. As shown in Fig. 1(g), the stable energy evolution means that α-$C_{60}$- 2D does not undergo bond breaking and generation in the dynamical evolution during the simulation, *i.e.*, energy conservation is maintained in an isolated environment. In addition, the embedding in Fig. 1(g) that describes the structural bonding characteristics shows that the bridge bonds between fullerene clusters can still remain intact, implying that α-$C_{60}$-2D is thermally stable. Likewise, the thermal stability of β-$C_{60}$-2D and γ-$C_{60}$-2D is also verified as shown in Supplementary Figure S1. Further, the mechanical stability of α-$C_{60}$-2D is confirmed by elastic constants $C_{ij}$ (*i, j* = 1, 2, 6) with the standard Voigt notation of 1-*xx*, 2-*yy*, and 6-*xy*. As shown in Fig. 1(h), $C_{11}$, $C_{12}$, $C_{16}$, $C_{22}$ and $C_{66}$ values of 186.83, 24.08, -1.25, 136.12, 62.20 N/m obey the Bonn-Huang criterion, *i.e.* $C_{11}C_{22} - C_{12}^2 > 0$, $C_{11} > 0$ and det($C_{ij}$) > 0, implying its mechanical stability[34].

The covalent bridge bonds in α-$C_{60}$-2D connect together the fullerene clusters in *xy* plane, which can be quantified by electron localization functions (ELF). The ELF is defined as $(1 + \{K(r)/K_h[\rho(r)]\}^2)^{-1}$, where $K_h[\rho(r)]$ is the curvature in a homogeneous (*h*) electron gas of density $\rho$ in position *r*[35]. The magnitude of the ELF indicates the degree of electron localization, which is limited to the region of 0 (completely delocalized) to 1 (completely localized). As shown in Fig. 1(i), similar to the degree of bonding in fullerenes, a highly localized electron density is found around intercluster carbon atoms, showing that strong covalent bonds are formed. Note that the electron density distribution over intercluster bonds exhibits strong in-plane asymmetry, which could be responsible for the anisotropic physical properties as described below.

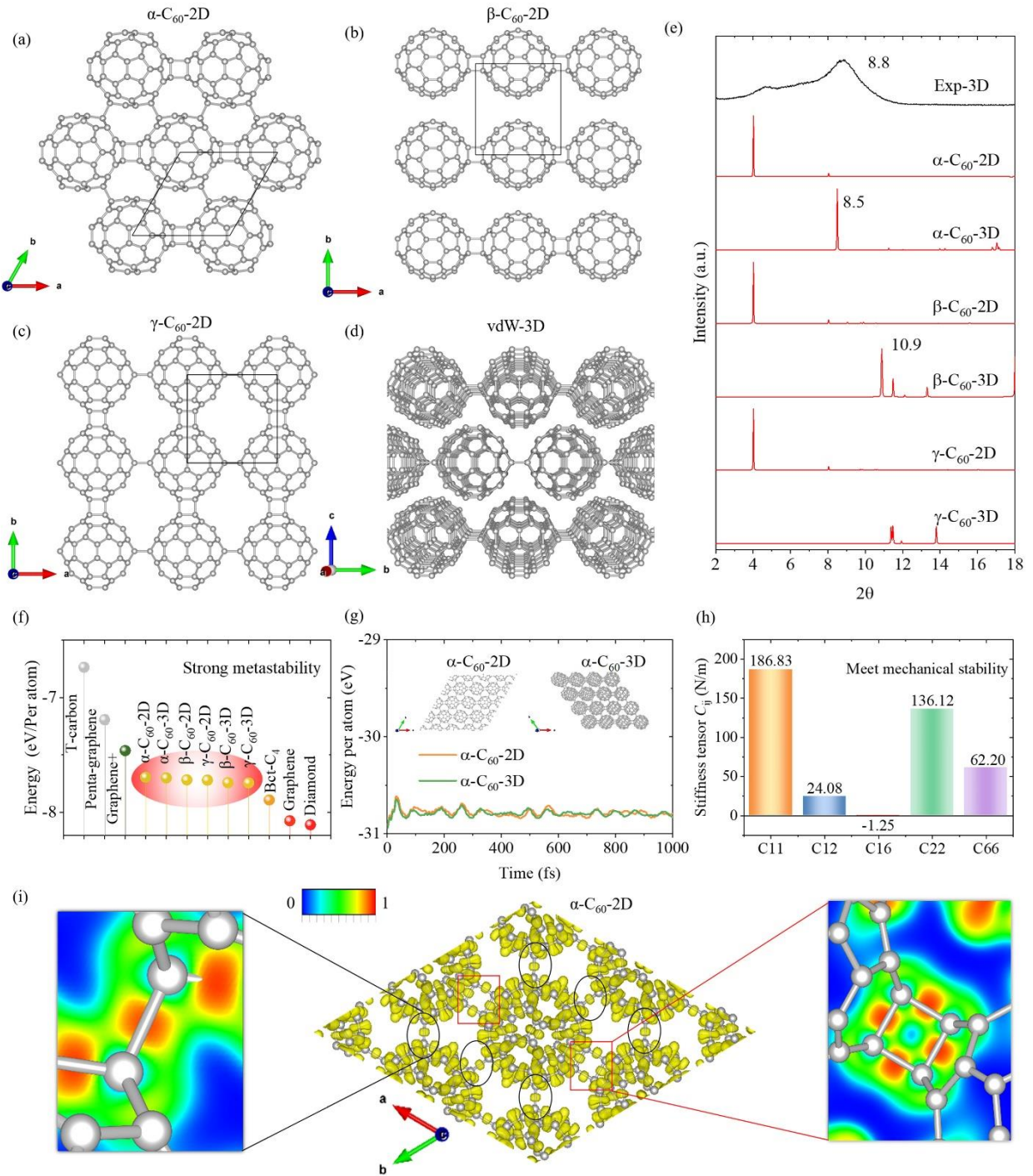

Figure 1. Two-dimensional fullerenes with different assembly forms and corresponding van der Waals (vdW) crystals. The crystal structures of (a) $\alpha$-$C_{60}$-2D, (b) $\beta$-$C_{60}$-2D, (c) $\gamma$-$C_{60}$-2D, and (d) vdW-3D. (e) The simulated and experimental measured X-ray diffraction of fullerene networks; (f) Comparison of the average energy of different carbon allotropes. (g) The evolution of energy in the AIMD simulations of α-$C_{60}$-2D and α-$C_{60}$-3D. (h) The elastic constants of α-$C_{60}$-2D. (i) Insight from 3D and 2D electron localized function (ELF) into the bridge bonds in α-$C_{60}$-2D.

**Flat band and dual CBM**-. The electronic properties of $C_{60}$ polymers are comprehensively investigated. Figs. 2(a, b) show the band structures and density of states (DOS) of the monolayer and the corresponding 3D vdW-type α-$C_{60}$ polymer. It is well known that graphene possesses the Dirac cone caused by the linear contact of $p_z$ orbitals at the valence band maximum (VBM) and the conduction band minimum (CBM). Compared with the independent $p_z$ orbitals in graphene, the $p_z$ orbitals of 2D and 3D $C_{60}$ polymers exhibit strong hybridization with $p_x/p_y$ orbitals due to bridged $sp^3$ bonding [Fig. 1(a-d)], suggesting that $sp^3$ hybridization plays a critical role. Interestingly, the VBM and $CBM_1$ of α-$C_{60}$-2D are located at the high symmetry points of Γ and Y, respectively, revealing the semiconducting nature with an indirect band gap as shown in Fig. 2(a). Notably, the energy band where the CBM locates has a weak dependence on the wave vector, indicating a flat band, which can be visually revealed through the sharp DOS peak at the CBM. Therefore, the α-$C_{60}$-2D exhibits dual CBMs, where the energy level difference between Y ($CBM_1$) and Γ ($CBM_2$) at the high symmetry point is almost negligible compared to the band gap and computational uncertainty, *i.e.*, ~0.01 eV. With the influence of vdW interactions, the flat band causes the band gap to easily switch between direct and indirect forms. As shown in Fig. 2(b), the $CBM_1$ of α-$C_{60}$-3D is located at Γ point and $CBM_2$ is located at Y point, behaving conversely with α-$C_{60}$-2D. Intriguingly, such a transition is also found to occur in the monolayer and vdW polymers of β-$C_{60}$ and γ-$C_{60}$ as shown in Figs. 2(c-f) (More details can be found in the Supplemental Figure S2 and S3.).

**Band gap and direct semiconductors** -. To capture the accurate band gap of α-$C_{60}$-2D, the HSE06 functional is further performed as shown in Fig. 2(g). A band gap of 1.44 eV is found in α-$C_{60}$-2D, which is relatively consistent with the experimental value of 1.66 eV, revealing the mid-bandgap semiconductor properties. Due to the flat band, the dual CBMs can be well observed at points Y ($CBM_1$) and Γ ($CBM_2$) in both the PBE and HSE06 functionals. The weak difference between $CBM_1$ and $CBM_2$ from 0.01 to 0.05 eV indicates the robustness of the flat band and dual CBM features. Thus, direct bandgap properties of 1.49 eV can be observed in α-$C_{60}$-2D. Generally, the carbon allotropes possess band gaps in excess of 2 eV and tend to exhibit insulating properties, such as diamond, V-carbon, W-carbon, M-carbon, Cco-$C_8$, Bct-$C_4$, T-carbon, and penta-graphene[13,36–39]. Compared to the typical carbon allotropes with wider band gaps, the fullerenes connected by bridging covalent bonds as studied in this work form a semiconductor α-$C_{60}$-2D with a moderate band gap. Both experimental and theoretical results reveal that the band gap of α-C$_{60}$-2D is not far from the bulk silicon[40]. Such a band gap could make α-C$_{60}$-2D possess promising potential applications in micro- and nano-electronic devices.

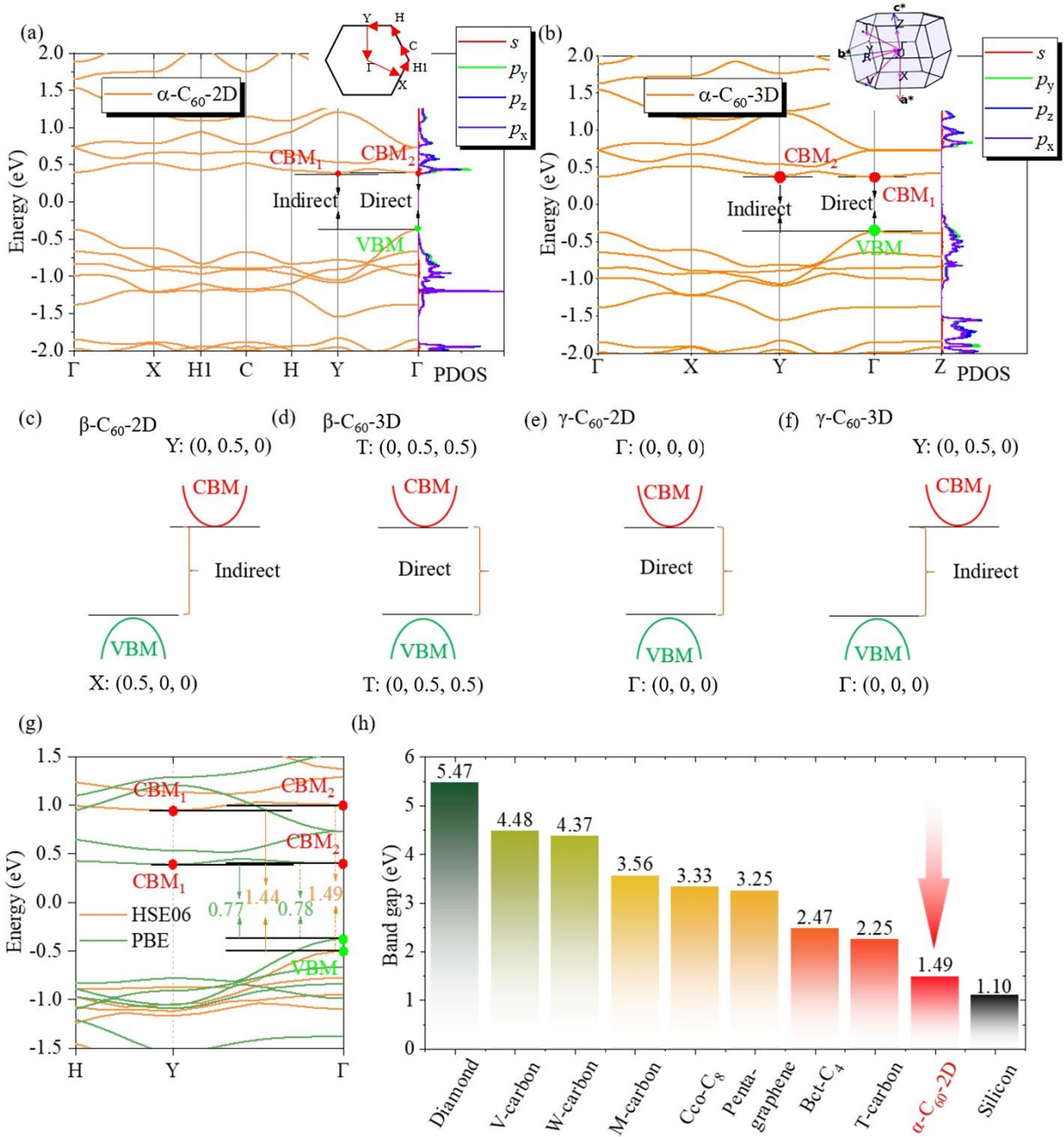

Figure 2. Electronic structures of C$_{60}$ monolayers. The band structures and projected density of states of (a) α-C$_{60}$-2D and (b) α-C$_{60}$-3D. Van der Waals tunable band gaps for (c) β-C$_{60}$-2D, (d) β-C$_{60}$-3D, (e) γ-C$_{60}$-2D, and (f) γ-C$_{60}$-3D. (g) The electronic energy band structures of α-C$_{60}$-2D calculated using HSE06 and PBE functionals. (h) Band gaps for different carbon allotropes, bulk silicon, and α-C$_{60}$-2D.

**Optical linear dichotomy** -. Based on the accurate band gap, the optical properties of α-C$_{60}$-2D are explored. The dielectric function of α-C$_{60}$-2D can be obtained, where the frequency-dependent dielectric function with the real $\varepsilon_1(\omega)$ and imaginary $\varepsilon_2(\omega)$ parts of α-C$_{60}$-2D can be obtained based on the Kramers-Kronig dispersion relation[41]. For 2D monolayers, we only consider the in-plane optical absorption. The anisotropy in fullerene network is significant due to the asymmetric bridging bond arrangement. As shown in Fig. 3(a, b), the optical absorption exhibits strong anisotropy in *xy* plane, where the optical absorption along the *xy* direction is significantly lower than that in the *xx* (and *yy*) direction. Such a phenomenon reveals a strong linear dichroism in α-C$_{60}$-2D. Linear dichroism allows the orientation of α-C$_{60}$-2D single-crystal samples to be determined by spectroscopy, and the feasible experimental setup can be implemented as shown in Fig. 3(a). When incident light at a specific angle and near normal incidence, the absorption spectrum will produce distinct peaks and valleys as the sample is rotated due to anisotropy, especially in *xx* (*yy*) and *xy* directions.

**Optical absorption** -. Different absorption intensities are found in visible region (rainbow zone) as shown in Fig. 3(b). The absorption peak in *yy* direction is close to $3.5 \times 10^5$ cm$^{-1}$ when the wavelength is in the range of 400-450 nm, which is stronger than that in *xx* direction. Such strong peaks reveal strong visible light absorption capabilities for α-C$_{60}$-2D. The ultraviolet (UV) region is mainly divided into three regions, *i.e.*, short-wave ultraviolet (UVC: 200-280nm), medium-wave ultraviolet (UVB: 280-320nm) and long-wave ultraviolet (UVA: 320-400nm). UV radiation from tanned and burned skin comes primarily from the UVA and UVB regions while UVC radiation is usually blocked by the ozone layer. The optical absorption in *xx* and *yy* directions in the UVA region reaches $5 \times 10^5$ cm$^{-1}$. Such a strong absorption peak in *yy* direction is maintained at $4 \times 10^5$ cm$^{-1}$ even in the UVB region. The strong absorption peaks of α-C$_{60}$-2D in UVA and UCB indicate that it can be utilized as wearable devices material to protect UV radiation. In addition, the results of the PBE functional are also supplied in Supplementary Figure S4 for reference.

**Mechanical properties** -. Based on the elastic modulus, Young's modulus *E*, shear modulus *G* and Poisson's ratio *v* of α-C$_{60}$-2D are obtained as shown in Figs. 3(c-e). The values of *E*, *G* and *v* in different directions are 133-183 N/m, 61-67 N/m, and 0.13-0.20, respectively, and the anisotropic ratio can reach 1.4, 1.1, and 1.6, respectively. The maximum and minimum *E* of α-C$_{60}$-2D emerge in the orthogonal direction, being much lower than that of graphene (329 N/m) and penta-graphene (271 N/m), which is at the same level as graphene+ (120 N/m). In contrast, the maximum and minimum values of *G* and *v* are separated by ~30°, respectively. The low Young's modulus originates from the less dense bridging bonds, indicating that α-$C_{60}$-2D is more prone to deformation when external perturbations are imposed, which provides the opportunity to modulate the mechanical performance of α-$C_{60}$-2D with strain engineering.

**Lithium ion batteries -.** Among carbon-based materials, the layered vdW structure of graphite exhibits a large interlayer accommodation for lithium (Li) ions, serving as an anode material for lithium ion batteries (LIBs).[42] Due to the spherical cage-like structure, α-$C_{60}$-2D is expected to exhibit promising energy storage potential capability. Fig. 3(f) shows the migration paths of Li atoms along different directions. As shown in Fig. 3(g), the anisotropic energy evolution is found during the migration of Li atoms in α-$C_{60}$-2D, where the X path has the smallest migration barrier for Li atoms, indicating that Li migrates more easily along the X path. The low barrier on the X path indicates stronger Li-ion diffusivity than along the Y path[43], promising the anisotropic potential applications in lithium batteries of α-$C_{60}$-2D.

**Carrier mobility and Seebeck coefficient** -. Considering that α-$C_{60}$-2D is a semiconductor with an intermediate bandgap, the electron-phonon coupling should not be dominated. Thus, we consider phonon scattering as the main factors to confine the carrier migration. Obviously, the electrical transport properties are anisotropic in α-$C_{60}$-2D. The effective mass m$^*$, elastic modulus $C_{2D}$, deformation potential $E_1$ in *x* direction are higher than those in *y* direction. In particular, the effective mass of electrons is significantly larger than that of holes due to the flat conduction band. Consequently, electrons (*x*: 0.056×10$^3$, *y*: 0.061×10$^3$ cm$^2$V$^{-1}$s$^{-1}$) tend to have lower carrier mobility compared to holes (*x*: 0.923×10$^3$, *y*: 0.912×10$^3$ cm$^2$V$^{-1}$s$^{-1}$), which is largely due to the large effective mass of the flat band near the CBM. The highest carrier mobility of 0.923×10$^3$ cm$^2$V$^{-1}$s$^{-1}$ is from holes in *x* direction, which is lower than the highest mobility of other 2D carbon allotropes such as graphene (3.289×10$^3$ cm$^2$V$^{-1}$s$^{-1}$)[44], and Tetrahexcarbon (14.63×10$^3$ cm$^2$V$^{-1}$s$^{-1}$)[45]. The low carrier mobility originates from the large effective mass during carrier migration due to the flat-band nature.

Based on the deformation potential theory, the electron relaxation time $\tau$ is calculated as listed in Table 1. The maximum relaxation time is caused by holes in $x$ direction, which is on the same order of magnitude with recently reported thermoelectric materials, such as $X_2Y_2Z_2$ type monolayers (100-160 fs)[46–48]. Moreover, Fig. 3(h) reveals a strong Seebeck coefficient at middle-low temperature. For instance, the Seebeck is around 3000 μV/K at 100 K, where the temperature is much lower than that of black phosphorus[49] and T-carbon[43]. The behavior of Seebeck coefficient indicates the possibility of fabricating low temperature thermoelectric devices with α-$C_{60}$-2D.

Table 1. Electrical transport properties of α-$C_{60}$-2D, including bandgap $E_g$, effective mass $m^*$, elastic modulus $C_{2D}$, deformation potential constant $E_1$, carrier mobility $\mu$, and relaxation time $\tau$.

| Direction | Carrier | $m^*$ ($m_0$) | $C_{2D}$ (J m$^{-2}$) | $E_1$ (eV) | $\mu$ ($10^3$ cm$^2$V$^{-1}$s$^{-1}$) | $\tau$ (fs) |
|---|---|---|---|---|---|---|
| $x$ | Electron | 1.22 | 167.08 | 6.88 | 0.056 | 38.72 |
|  | Hole | -0.60 |  | 3.63 | 0.923 | 315.12 |
| $y$ | Electron | 1.01 | 141.68 | 6.64 | 0.061 | 35.25 |
|  | Hole | -0.40 |  | 4.12 | 0.912 | 20.74 |

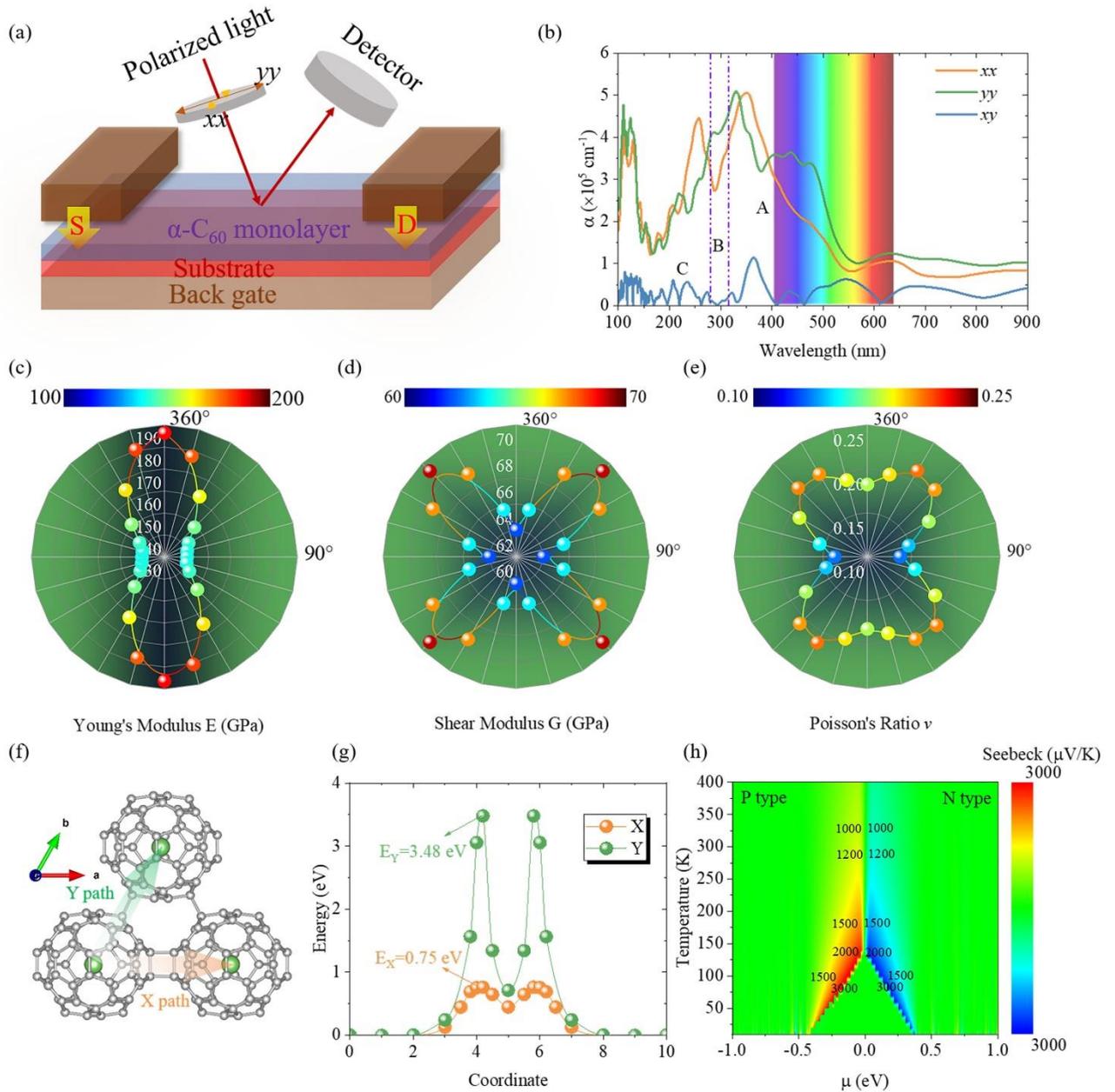

Figure 3. Anisotropic properties of α-$C_{60}$-2D. (a) Schematic depict of the experimental setup for the linear dichotomy, where the lattice orientation can be determined by spectral absorption. (b) Optical absorption of α-$C_{60}$-2D. (c) Young's modulus $E$, (d) shear modulus $G$, and (e) Poisson's ratio $v$ of α-$C_{60}$-2D. (f) Anisotropic lithium ion migration paths and the corresponding (g) energy evolution of lithium ions during migration in α-$C_{60}$-2D. (h) The Seebeck coefficients for α-$C_{60}$-2D.

## Conclusions

Based on the large-scale and comprehensive calculations, we have demonstrated that the experimentally synthesized novel fullerene polymer monolayers are 2D semiconductors with direct band gaps owing to the dual CBMs from the flat conduction band bottom. The calculated band gap of 1.49 eV is in good agreement with the experimental value of 1.66 eV. The structure of the experimentally synthesized fullerene polymer network is identified as α-$C_{60}$-2D. We uncover that strong covalent bonds are formed between intercluster carbon atoms. The asymmetric arrangement of covalent bridges between fullerene clusters leads to anisotropic optical, mechanical, and electrical transport properties. In particular, we find an optical linear dichroism in α-$C_{60}$-2D, which leads to anisotropic optical absorption allowing the determination of lattice orientation by spectroscopy. It is also revealed that α-$C_{60}$-2D has a strong optical absorption in the long-wave ultraviolet region and anisotropic mechanical features with low Young's modulus. The electron mobility of α-$C_{60}$-2D ($x$: $0.056 \times 10^3$, $y$: $0.061 \times 10^3$ cm$^2$V$^{-1}$s$^{-1}$) is found smaller than those of hole mobility ($x$: $0.923 \times 10^3$, $y$: $0.912 \times 10^3$ cm$^2$V$^{-1}$s$^{-1}$). In addition, Li ions are uncovered to migrate most easily along the X path in α-$C_{60}$-2D, showing the potential application in LIBs. The calculations also show that the Seebeck coefficient of α-$C_{60}$-2D can be as large as 3000 μV/K at 100 K, revealing its possible applications in thermoelectric nanodevices. The present study promises that α-$C_{60}$-2D could be a potential candidate for various potential applications in 2D electronic and optoelectronic devices.

## Methods

All the first-principles calculations are implemented by the *Vienna ab-initio simulation package*[50] (VASP) within the framework of density functional theory (DFT). The vacuum of 15 Å is selected to prevent interaction between layers for 2D system. The optB86b[51] functional was chosen to include the van der Waals interaction, which is the exchange correlation function based on projector augmented wave[52] (PAW) basis set with the kinetic energy cutoff of 700 eV . To obtain the relaxed configuration, an 5×5×1 Monkhorst-Pack[53] k-mesh is used to sample the Brillouin zone until the energy and Hellmann-Feynman force accuracy converge to 10$^{-6}$ eV and 10$^{-2}$ eV/Å, respectively. (More details can be found in the Supplemental Note S1 and Figure S5.).

## ACKNOWLEDGEMENTS

G.Q. is supported by the National Natural Science Foundation of China (Grant No. 52006057), the Fundamental Research Funds for the Central Universities (Grant Nos. 531119200237 and 541109010001), and the State Key Laboratory of Advanced Design and Manufacturing for Vehicle Body at Hunan University (Grant No. 52175013). G.S. is supported by the National Key R&D Program of China (Grant No. 2018YFA0305800), the Strategic Priority Research Program of CAS (Grant No. XDB28000000), the NSFC (Grant No. 11834014), and Platform Projects of Chinese Academy of Sciences. The numerical calculations have been done on the supercomputing system of the National Supercomputing Center in Changsha, the Shanghai advanced computing center.

**AUTHOR CONTRIBUTIONS**

G.S. and G.Q. conceived and supervised the project. L.Y. performed all the calculations and analysis, and the data in Fig. 2(g), Fig. 3(h) and Table 1 were calculated by J.X. All the authors contributed to interpreting the results. The paper was written by L.Y. with contributions from all the authors.

**DECLARATION OF INTERESTS**

The authors declare no competing interests

# Supplementary Materials for "Flat electronic band structure and anisotropic optical, mechanical, and thermoelectric properties of two-dimensional fullerene networks"

Linfeng Yu[1], Jinyuan Xu[1], Bo Peng[1], Guangzhao Qin[1,‡], and Gang Su[2,§]

[1]*State Key Laboratory of Advanced Design and Manufacturing for Vehicle Body, College of Mechanical and Vehicle Engineering, Hunan University, Changsha 410082, P. R. China*

[2]*Kavli Institute for Theoretical Sciences and CAS Center for Excellence in Topological Quantum Computation, University of Chinese Academy of Sciences, Beijing 100190, China*

*Correspondence:* gzqin@hnu.edu.cn, gsu@ucas.ac.cn

# NOTE 1. Computational Theory

Mechanically, the elastic constant matrix can be obtained according to Hooke's law[1]

$$\begin{bmatrix} \sigma_{xx} \\ \sigma_{yy} \\ \sigma_{xy} \end{bmatrix} = \begin{bmatrix} C_{11} & C_{12} & C_{16} \\ C_{12} & C_{22} & C_{26} \\ C_{61} & C_{62} & C_{66} \end{bmatrix} \begin{bmatrix} \varepsilon_{xx} \\ \varepsilon_{yy} \\ 2\varepsilon_{xy} \end{bmatrix} \qquad (1)$$

where $C_{ij}$ ($i,j$ = 1, 2, 6) is the stiffness tensor with the standard Voigt notation $ij$ (1-$xx$, 2-$yy$, and 6-$xy$). The $C_{ij}$ is calculated by the energy-strain method[2]

$$U(\varepsilon) = \frac{1}{2} C_{11} \varepsilon_{xx}^2 + \frac{1}{2} C_{22} \varepsilon_{yy}^2 + C_{12} \varepsilon_{xx} \varepsilon_{yy} + 2 C_{66} \varepsilon_{xy}^2 . \qquad (2)$$

To obtain the optical properties, the dielectric function with real $\varepsilon_1(\omega)$ and imaginary $\varepsilon_2(\omega)$ parts is written as [3]

$$\varepsilon(\omega) = \varepsilon_1(\omega) + i\varepsilon_2(\omega) . \qquad (3)$$

The imaginary part $\boldsymbol{\varepsilon_2(\omega)}$ is given by[4]

$$\varepsilon_2(\omega) = \frac{4\pi^2 e^2}{m^2 \pi^2} \times \sum_{C,V} |P_{C,V}|^2 \delta(E_C + E_V - \hbar\omega) , \qquad (4)$$

where $e$, m, $\omega$ are the electronic charge, the effective mass, and the angular frequency, respectively. $P_{C,V}$ represents the momentum conversion matrix between the conduction band $C$ and valence band $V$. $E$ is the electronic energy level. The delta function $\delta$ can enhance the energy conversion of electrons during the transition from band to band. $\varepsilon_2(\omega)$ should be normalized with the vacuum layers:[5]

$$\langle \varepsilon_2 \rangle = \frac{L}{d} \varepsilon_2 , \qquad (5)$$

where $L$ and $d$ are the layer spacing and effective thickness, respectively. Simultaneously, $\langle \varepsilon_1 \rangle$ can be expressed based on the Kramers-Kronig dispersion relation:[6]

$$\langle \varepsilon_1 \rangle = 1 + \frac{2}{\pi} \int_0^\infty \frac{\omega' \langle \varepsilon_2(\omega') \rangle}{\omega'^2 - \omega^2} d\omega' \quad . \tag{6}$$

Furthermore, the linear photon spectra are obtained based on the dielectric function. The absorption coefficient $\alpha(\omega)$ can be calculated as:[3]

$$\alpha(\omega) = \frac{\sqrt{2}\omega}{c} \left[ \frac{\sqrt{\varepsilon_1(\omega)^2 + \varepsilon_2(\omega)^2} - \varepsilon_1(\omega)}{2} \right]^{\frac{1}{2}} . \tag{7}$$

To calculate the electrical transport properties, the elastic constant is calculated by $C_{2D} = [\partial^2 E/\partial(\Delta x/x_0)^2]/S_0$ based on deformation potential (DP) theory, where $E$ is the system's total energy under a uniaxial strain along the $x$ axis. The DP constant is described as $E_1 = \partial E_{edge}/\partial(\Delta x/x_0)$, where $E_{edge}$ is the shift of the band edge. The effective mass $m^*$ can be obtained by $m^* = \hbar^2/(\partial^2 E/\partial k^2)$. Furthermore, the electron relaxation time $\tau$ can be calculated by[7]

$$\tau = \frac{\mu m^*}{e} = \frac{\hbar^3 C_{2D}}{k_B T m_d E_1^2} , \tag{8}$$

where $\hbar$ and $k_B$ are the Planck constant, Boltzmann constant, respectively. The absolute value of the Seebeck coefficient $S$ is calculated by[8]

$$S = \frac{8\pi^2 k_B^2}{3eh^2} m^* T \left(\frac{\pi}{3n}\right)^{2/3} . \tag{9}$$

Table S1. Lattice constants and electronic bandgaps of the 2D fullerene networks and the corresponding van der Waals crystals. The two values of bandgaps are for direct/indirect, respectively.

| Name | a (Å) | b (Å) | c (Å) | Bandgap (eV) | |
| --- | --- | --- | --- | --- | --- |
| | | | | PBE | HSE06 |
| α-$C_{60}$-2D | 9.14 | 9.07 | - | 0.78/0.77 | 1.49/1.44 |
| α-$C_{60}$-2D (bilayer) | 9.12 | 9.13 | - | 0.68/0.73 | - |
| α-$C_{60}$-3D | 9.13 | 9.06 | 11.31 | 0.74/0.76 | - |
| β-$C_{60}$-2D | 9.06 | 9.77 | - | 1.03/0.99 | - |
| β-$C_{60}$-3D | 9.07 | 9.79 | 14.61 | 0.60/0.79 | - |
| γ-$C_{60}$-2D | 9.09 | 8.99 | - | 1.11/1.17 | - |
| γ-$C_{60}$-3D | 9.13 | 9.03 | 14.83 | 0.87/0.75 | - |

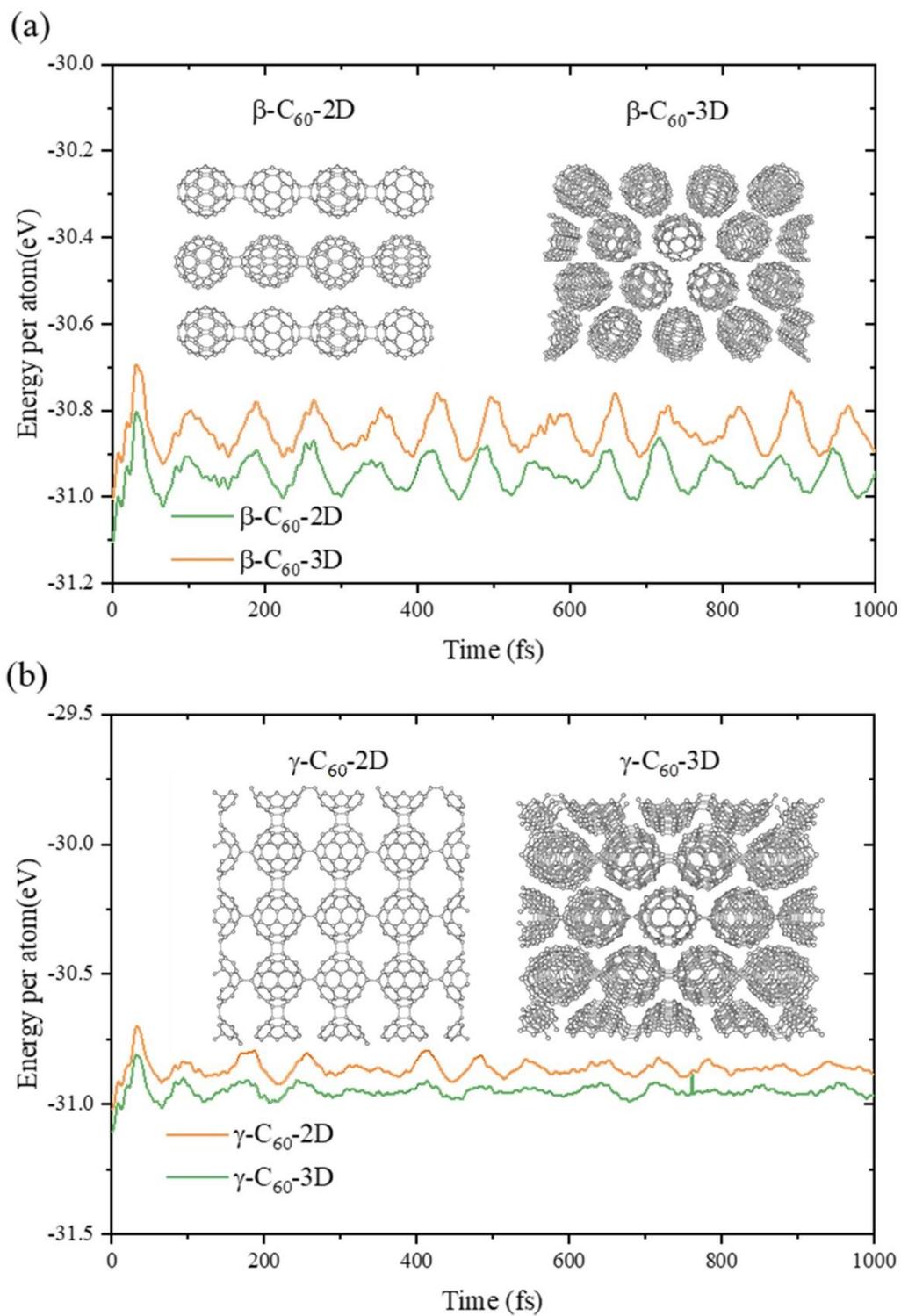

Figure S1: The evolution of energy with respect to time in AIMD simulations, where the atomic configuration at the final step of simulations are presented as insets.

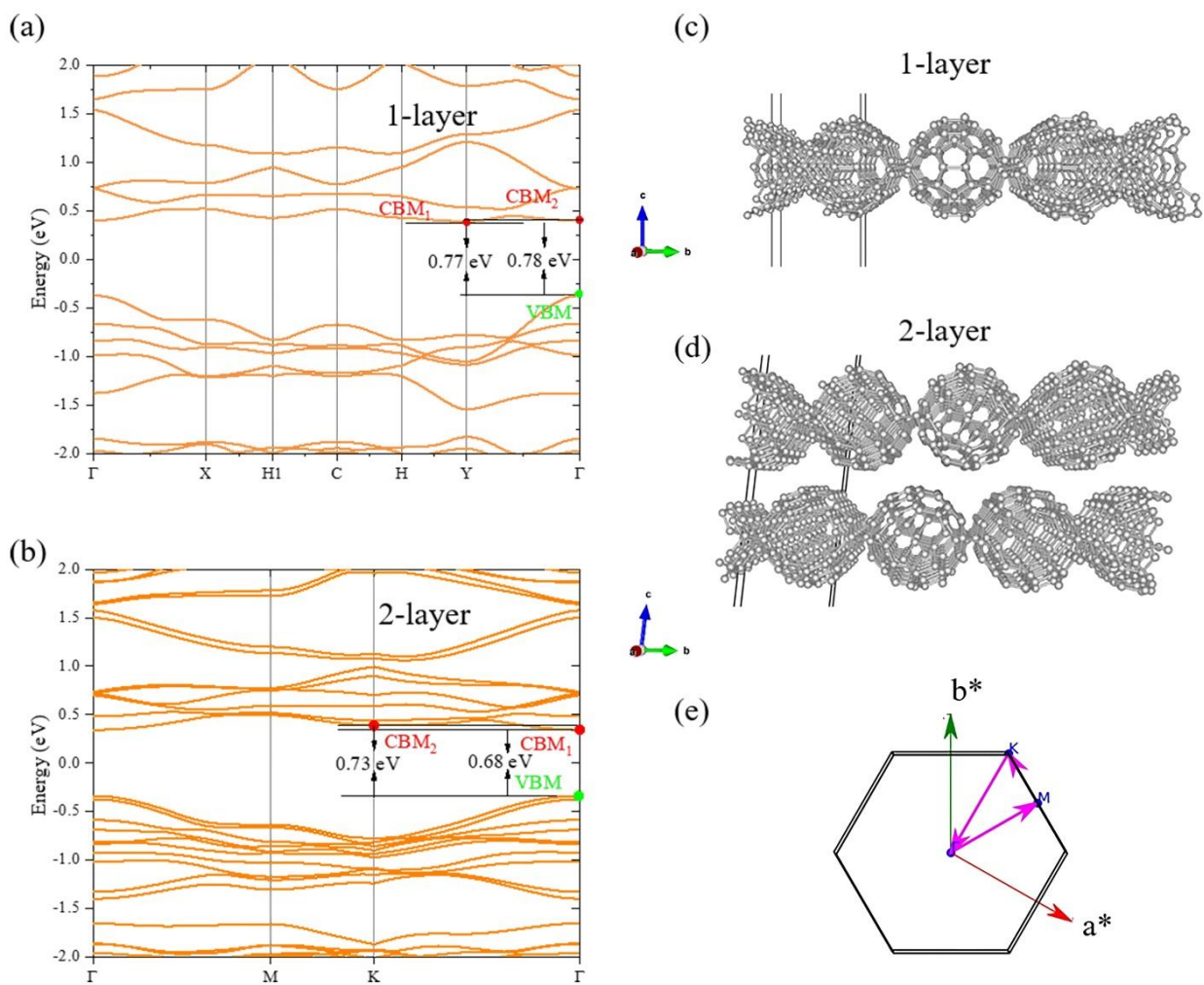

Figure S2. The effect of layer numbers on electronic structures. The band structures of (a) monolayer and (b) bilayer α-$C_{60}$. The unit cell used for the band structure calculation of (c) monolayer and (d) bilayer α-$C_{60}$; (d) Brillouin zone of bilayer α-$C_{60}$ for band structure calculations.

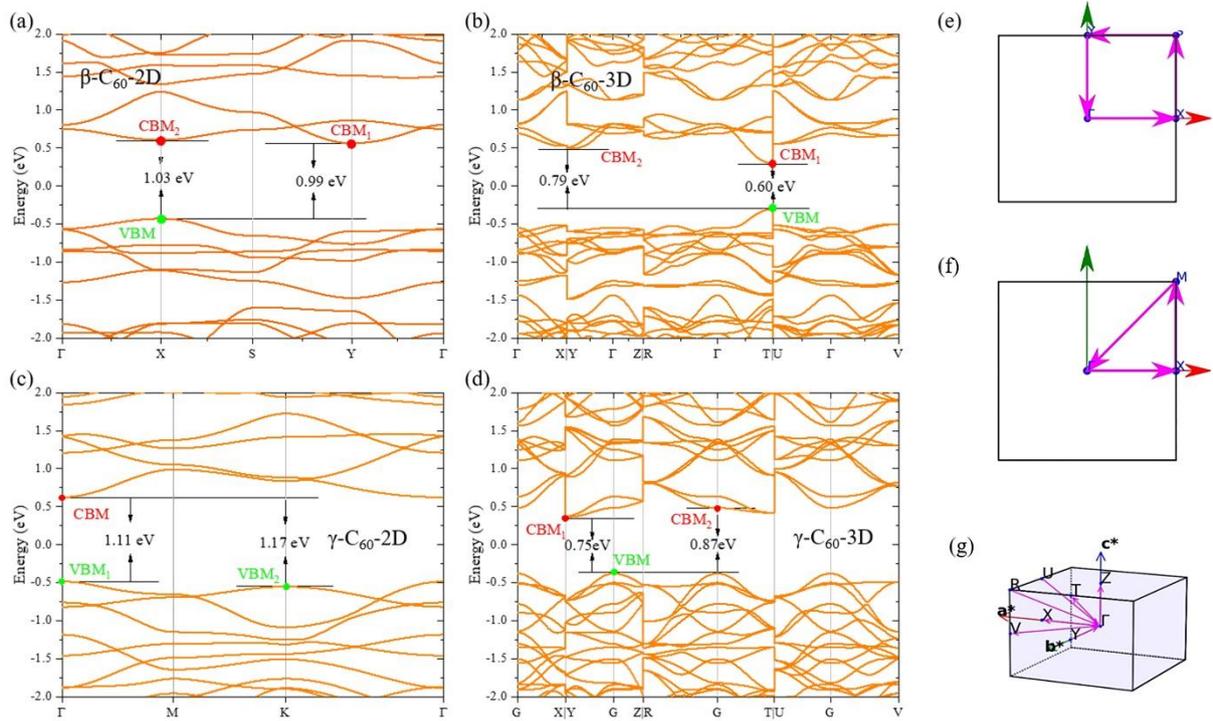

Figure S3. Band structure of (a) β-C$_{60}$-2D, (b) γ-C$_{60}$-2D, (c) β-C$_{60}$-3D and (d) γ-C$_{60}$-3D; Brillouin zone of (e) β-C$_{60}$-2D, (f) γ-C$_{60}$-2D, (g) β-C$_{60}$-3D and γ-C$_{60}$-3D.
25

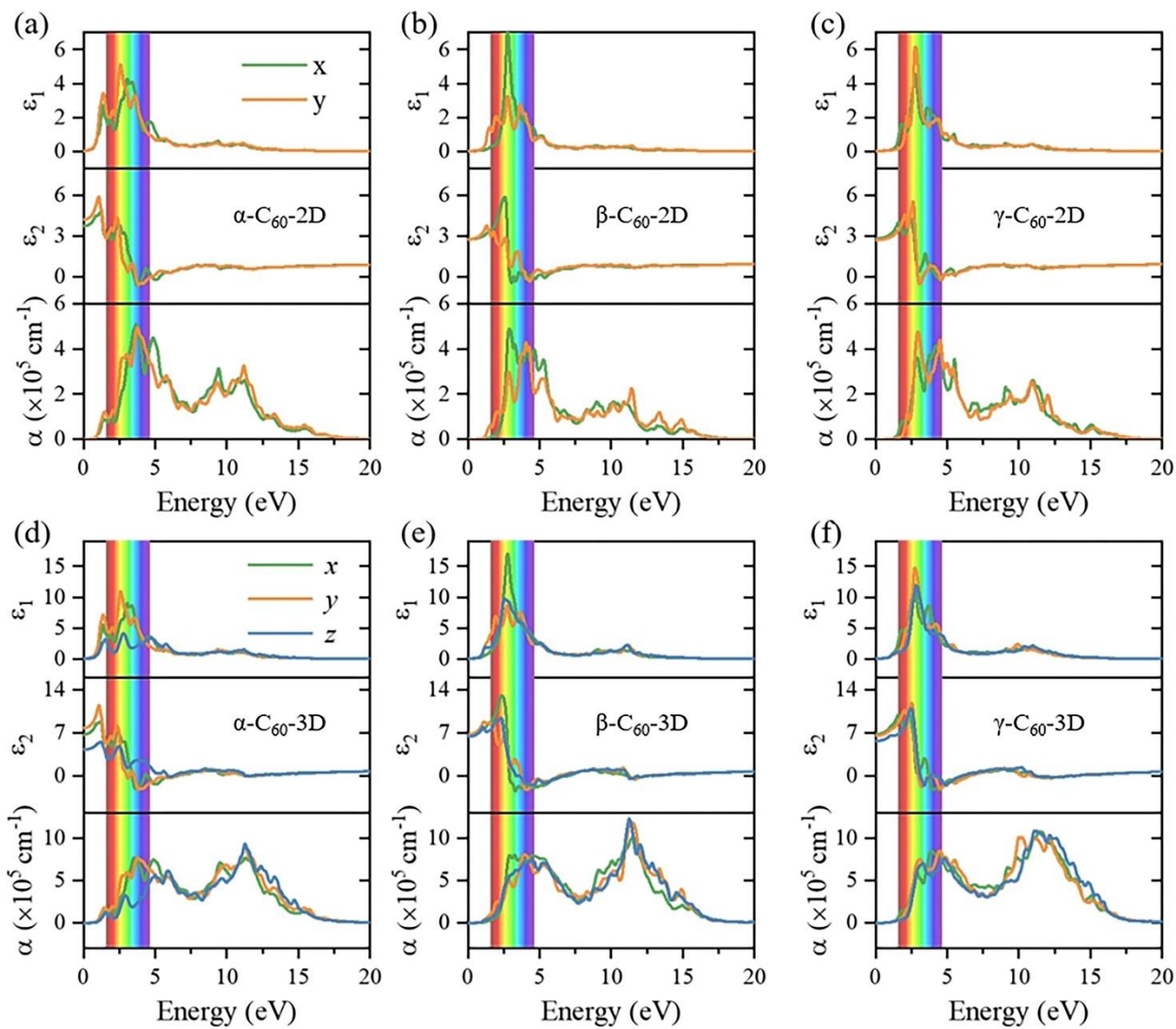

Figure S4. Optical properties of two-dimensional fullerene networks and corresponding 3D van der Waals crystals for (a) α-$C_{60}$-2D; (b) β-$C_{60}$-2D; (c) γ-$C_{60}$-2D; (d) α-$C_{60}$-3D, (e) β-$C_{60}$-3D; (f) γ-$C_{60}$-3D.

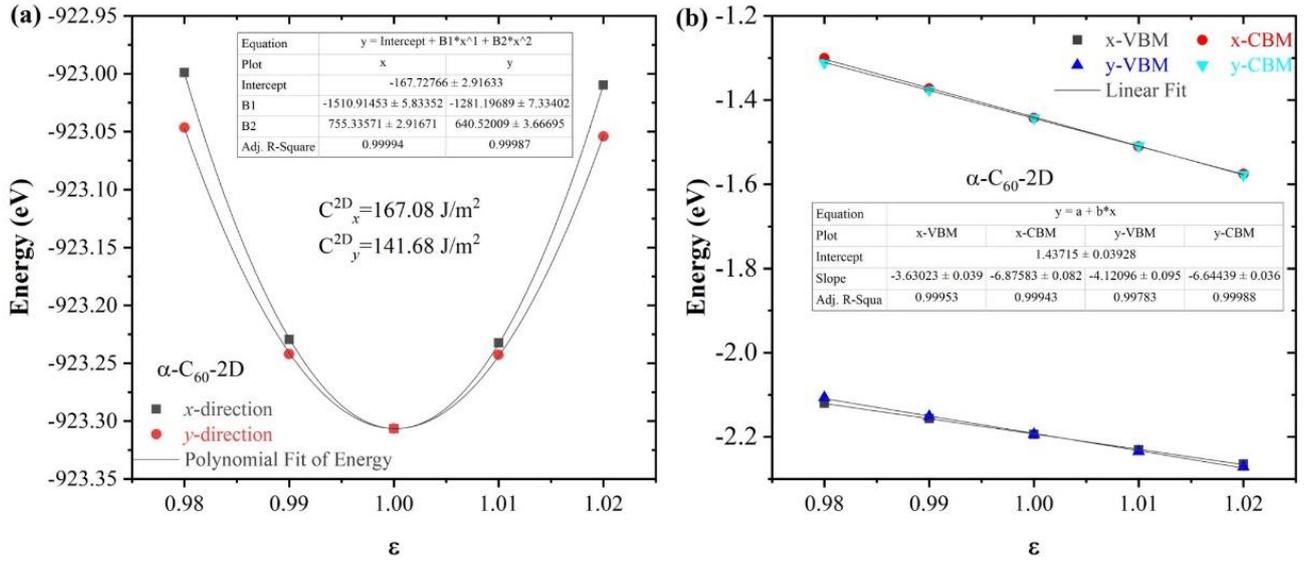

Figure S5: Calculated deformation potential parameters and the fitting information. (a) The variation of energy under strain along the *x*-/*y*-directions for the 2D *in-plane* elastic modulus ($C^{2D}$). (b) The CBM and VBM along the *x*-/*y*-directions as a function of strain for the deformation potential constants ($E_1$). Noted that the effective mass was calculated by *vaspkit* [9] code.